\documentclass[aps,prl,twocolumn,superscriptaddress,preprintnumbers,amsmath,amssymb]{revtex4}
\usepackage{graphicx}
\usepackage{dcolumn}
\usepackage{bm}
\usepackage{color}
\usepackage{adjustbox}
\usepackage{multirow}
\usepackage{ulem}
\usepackage{ragged2e}

\begin{document}

\title{Fourfold Differential Photoelectron Circular Dichroism}

\author{K.~Fehre}\email{fehre@atom.uni-frankfurt.de}
\affiliation{Institut f\"{u}r Kernphysik, Goethe-Universit\"{a}t, Max-von-Laue-Strasse 1, 60438 Frankfurt am Main, Germany}

\author{N.~M.~Novikovskiy}
\affiliation{Institut\,f\"{u}r\,Physik\,und\,CINSaT,\,Universit\"{a}t\,Kassel,\,Heinrich-Plett-Strasse\,40,\,34132\,Kassel,\,Germany}
\affiliation{Institute of Physics, Southern Federal University, 344090 Rostov-on-Don, Russia}

\author{S.~Grundmann}
\affiliation{Institut f\"{u}r Kernphysik, Goethe-Universit\"{a}t, Max-von-Laue-Strasse 1, 60438 Frankfurt am Main, Germany}

\author{G.~Kastirke}
\affiliation{Institut f\"{u}r Kernphysik, Goethe-Universit\"{a}t, Max-von-Laue-Strasse 1, 60438 Frankfurt am Main, Germany}

\author{S.~Eckart}
\affiliation{Institut f\"{u}r Kernphysik, Goethe-Universit\"{a}t, Max-von-Laue-Strasse 1, 60438 Frankfurt am Main, Germany}

\author{F.~Trinter}
\affiliation{Institut f\"{u}r Kernphysik, Goethe-Universit\"{a}t, Max-von-Laue-Strasse 1, 60438 Frankfurt am Main, Germany}
\affiliation{Molecular Physics, Fritz-Haber-Institut der Max-Planck-Gesellschaft, Faradayweg 4-6, 14195 Berlin, Germany}

\author{J.~Rist}
\affiliation{Institut f\"{u}r Kernphysik, Goethe-Universit\"{a}t, Max-von-Laue-Strasse 1, 60438 Frankfurt am Main, Germany}

\author{A.~Hartung}
\affiliation{Institut f\"{u}r Kernphysik, Goethe-Universit\"{a}t, Max-von-Laue-Strasse 1, 60438 Frankfurt am Main, Germany}

\author{D.~Trabert}
\affiliation{Institut f\"{u}r Kernphysik, Goethe-Universit\"{a}t, Max-von-Laue-Strasse 1, 60438 Frankfurt am Main, Germany}

\author{C.~Janke}
\affiliation{Institut f\"{u}r Kernphysik, Goethe-Universit\"{a}t, Max-von-Laue-Strasse 1, 60438 Frankfurt am Main, Germany}

\author{G.~Nalin}
\affiliation{Institut f\"{u}r Kernphysik, Goethe-Universit\"{a}t, Max-von-Laue-Strasse 1, 60438 Frankfurt am Main, Germany}

\author{M.~Pitzer}
\affiliation{Institut f\"{u}r Kernphysik, Goethe-Universit\"{a}t, Max-von-Laue-Strasse 1, 60438 Frankfurt am Main, Germany}

\author{S.~Zeller}
\affiliation{Institut f\"{u}r Kernphysik, Goethe-Universit\"{a}t, Max-von-Laue-Strasse 1, 60438 Frankfurt am Main, Germany}

\author{F.~Wiegandt}
\affiliation{Institut f\"{u}r Kernphysik, Goethe-Universit\"{a}t, Max-von-Laue-Strasse 1, 60438 Frankfurt am Main, Germany}

\author{M.~Weller}
\affiliation{Institut f\"{u}r Kernphysik, Goethe-Universit\"{a}t, Max-von-Laue-Strasse 1, 60438 Frankfurt am Main, Germany}

\author{M.~Kircher}
\affiliation{Institut f\"{u}r Kernphysik, Goethe-Universit\"{a}t, Max-von-Laue-Strasse 1, 60438 Frankfurt am Main, Germany}

\author{M.~Hofmann}
\affiliation{Institut f\"{u}r Kernphysik, Goethe-Universit\"{a}t, Max-von-Laue-Strasse 1, 60438 Frankfurt am Main, Germany}

\author{L.~Ph.~H.~Schmidt}
\affiliation{Institut f\"{u}r Kernphysik, Goethe-Universit\"{a}t, Max-von-Laue-Strasse 1, 60438 Frankfurt am Main, Germany}

\author{A.~Knie}
\affiliation{Institut\,f\"{u}r\,Physik\,und\,CINSaT,\,Universit\"{a}t\,Kassel,\,Heinrich-Plett-Strasse\,40,\,34132\,Kassel,\,Germany}

\author{A.~Hans}
\affiliation{Institut\,f\"{u}r\,Physik\,und\,CINSaT,\,Universit\"{a}t\,Kassel,\,Heinrich-Plett-Strasse\,40,\,34132\,Kassel,\,Germany}

\author{L.~Ben~Ltaief}
\affiliation{Department of Physics and Astronomy, Aarhus University, 8000 {\AA}rhus, Denmark}

\author{A.~Ehresmann}
\affiliation{Institut\,f\"{u}r\,Physik\,und\,CINSaT,\,Universit\"{a}t\,Kassel,\,Heinrich-Plett-Strasse\,40,\,34132\,Kassel,\,Germany}

\author{R.~Berger}
\affiliation{Theoretical Chemistry, Universit\"{a}t Marburg, Hans-Meerwein-Strasse 4, 35032 Marburg, Germany}

\author{H.~Fukuzawa}
\affiliation{Institute of multidisciplinary research for advanced materials, Tohoku University, Sendai 980-8577, Japan}

\author{K.~Ueda}
\affiliation{Institute of multidisciplinary research for advanced materials, Tohoku University, Sendai 980-8577, Japan}

\author{H.~Schmidt-B\"ocking}
\affiliation{Institut f\"{u}r Kernphysik, Goethe-Universit\"{a}t, Max-von-Laue-Strasse 1, 60438 Frankfurt am Main, Germany}

\author{J.~B.~ Williams}
\affiliation{Department of Physics, University of Nevada, Reno, Nevada 89557, USA}

\author{T.~Jahnke}
\affiliation{European XFEL, Holzkoppel 4, 22869 Schenefeld, Germany}

\author{R.~D\"orner}
\affiliation{Institut f\"{u}r Kernphysik, Goethe-Universit\"{a}t, Max-von-Laue-Strasse 1, 60438 Frankfurt am Main, Germany}

\author{M.~S.~Sch\"offler}\email{schoeffler@atom.uni-frankfurt.de}
\affiliation{Institut f\"{u}r Kernphysik, Goethe-Universit\"{a}t, Max-von-Laue-Strasse 1, 60438 Frankfurt am Main, Germany}

\author{Ph.~V.~Demekhin}\email{demekhin@physik.uni-kassel.de}
\affiliation{Institut\,f\"{u}r\,Physik\,und\,CINSaT,\,Universit\"{a}t\,Kassel,\,Heinrich-Plett-Strasse\,40,\,34132\,Kassel,\,Germany}

\date{\today}

\begin{abstract}
We report on a joint experimental and theoretical study of photoelectron circular dichroism (PECD) in methyloxirane. By detecting O 1s-photoelectrons in coincidence with fragment ions, we deduce the molecule's orientation and photoelectron emission direction in the laboratory frame. Thereby, we retrieve a fourfold differential PECD clearly beyond 50\%. This strong chiral asymmetry is reproduced by \textit{ab initio} electronic structure calculations. Providing such a pronounced contrast makes PECD of fixed-in-space chiral molecules an even more sensitive tool for chiral recognition in the gas phase.
\end{abstract}

\maketitle
The interaction of chiral matter with circularly polarized light depends on the light's helicity, giving rise to so-called \textit{chiroptical} effects. Scalar observables, such as circular dichroism (CD, \cite{CD1}), rely on the interference between the electric-dipole and magnetic-dipole light-matter interactions. Accordingly, the magnitude of the effect is typically very small, i.e., on the order of 10$^{-6}$--10$^{-3}$ of the total absorption. Because of such rather weak contrast, this chiroptical effect is routinely utilized for chiral recognition mainly in the condensed phase \cite{CD2}. A pioneering theoretical prediction \cite{PECD1} of a substantial forward-backward asymmetry in the laboratory-frame angular distributions of photoelectrons emitted from randomly oriented chiral molecules (and a later experimental verification of this effect  \cite{PECD2}) paved the way for establishing a new tool for chiral recognition in the gas phase. Because this effect, known as photoelectron circular dichroism (PECD, \cite{REV1,REV2,REV3}), relies purely on the electric-dipole light-matter interaction, it is much stronger than the \textit{conventional} CD. Its strength typically reaches a few percent. PECD is a universal chiroptical effect \cite{Beaulieu16NJP} persisting in many regimes: one-photon ionization \cite{REV1,REV2,REV3}, resonance-enhanced multiphoton ionization \cite{Lux12AngChm,Lehmann13jcp}, above-threshold ionization  \cite{Beaulieu16NJP,Lux16ATI}, strong-field ionization \cite{Beaulieu16NJP,Fehre19}, and multiphoton ionization by bichromatic fields \cite{Beaulieu17as,Beaulieu18PXCD,PRLw2w,Rozen19}. Because of its high contrast as compared to the conventional CD, PECD became a well-established tool for  enantiomeric excess determination in the gas phase  with a sub-percent resolution \cite{Kastner16ee}.

Obviously, further improving the sensitivity of the enantiomeric excess determination by means of PECD would require a considerable increase of its strength. But what are the limits of PECD and how can it be enhanced \cite{Olga}? One of the discussed routes is to utilize a coherent control scheme to optimize the light properties, which may increase the chiral response up to about 68\% \cite{Christiane} (which corresponds to about 34\% for the normalized difference discussed in this work). Alternatively, since PECD of randomly oriented molecules is just a small fraction of a non-averaged chiral response \cite{PECD1} and, thus, the averaging over different molecular orientations inevitably leads to a loss of information, it is a viable route to increase the observed effect by fixing the target molecules in space. For diatomic (and thus achiral) molecules, this strategy yields up to 100\% circular dichroism in the angular distribution (CDAD, \cite{CDAD1,CDAD2}). The CDAD effect is, however, confined in the polarization plane of the circularly polarized light (i.e., it occurs without establishing a forward-backward asymmetry as in the PECD \cite{Pier20}). Our theoretical predictions \cite{Tia17} illustrate that fixing the molecular orientation in all three spatial dimensions may also result in a PECD of 100\%. In this previous study \cite{Tia17} on methyloxirane (C$_3$H$_6$O) and in the subsequent study \cite{Nalin21} on the closely related trifluoromethyloxirane  (C$_3$H$_3$F$_3$O) molecules, we made experimentally only the first step towards this ultimate goal: By fixing just one molecular-orientation axis in space, we were able to demonstrate an enhancement of the PECD of uniaxially oriented chiral molecules up to about 10--20\%.

Recently, we made the second step towards achieving the PECD of fixed-in-space chiral molecules  \cite{Fehre21}: We measured a polarization-averaged molecular-frame photoelectron angular distribution (MFPAD, \cite{MFPAD1,MFPAD2}) of methyloxirane. Being emitted from a site- and element-specific core level, a photoelectron wave serves as a messenger of the molecular structure, providing the information on the target and on the dynamics of the process. Using these polarization-averaged MFPADs, we achieved an enantiosensitive molecular structure determination with a bond-length resolution of about 5\% for strong and even weak atomic scatterers alike \cite{Fehre21}. However, such MFPADs are insensitive to the polarization state of the ionizing light, since all directions from which the light impinges onto the molecule are summed over. In the present work, we now exploit the sensitivity of photoelectron angular emission distributions to the helicity of circularly polarized light and, thus, make the final step to measure the differential PECD of a fixed-in-space methyloxirane molecule. Thereby, we demonstrate experimentally and theoretically that the normalized chiral asymmetry obtained as a function of two photoelectron emission angles and two molecular orientation angles can indeed exceed 50\%.

The experimental data were recorded at beamline SEXTANTS of synchrotron SOLEIL, Saint-Aubin, France and the experimental setup and conditions equal those reported in our previous works on methyloxirane \cite{Tia17,Fehre21}. The photoelectron energy of 550~eV was set to create O~1s-photoelectrons with a kinetic energy of $11.5\pm1.5$~eV. The coincident detection of the momenta of charged ionic fragments and electrons was carried out by using a COLTRIMS reaction microscope \cite{COLTRIMS1,COLTRIMS2}. We recorded the full data set for R- and S-enantiomers, as well as for light with positive and negative helicities (LCP and RCP), respectively. To ensure the same experimental conditions and minimize systematical errors, we toggled between LCP and RCP every two hours. To increase statistics, we combined  results obtained for the two light helicities in the data analysis by applying the symmetry rules discussed below. We have checked the consistency of our data by comparing the results gathered for R- and S-methyloxirane. For the sake of brevity, we present here only the S-methyloxirane results.

\begin{figure}
\centering
\includegraphics[width=0.45\textwidth]{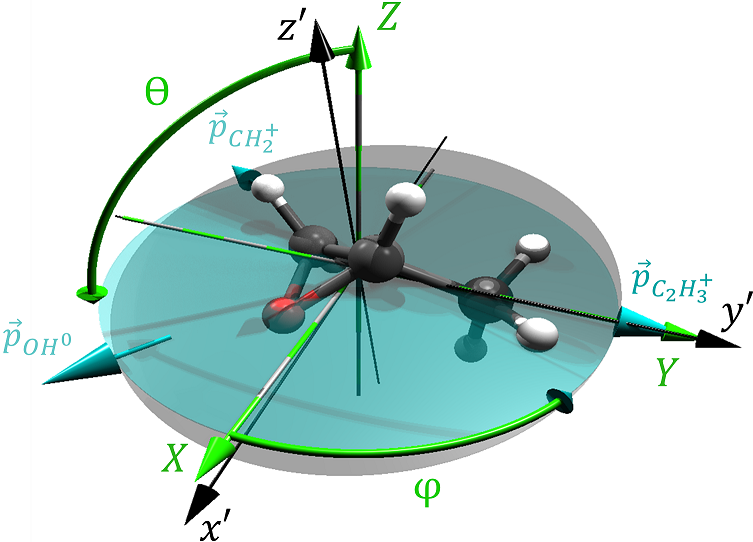}
\caption{Definition of the \textit{fragment} coordinate system (green axes $X,Y$, and $Z$) for S-methyloxirane with the help of the asymptotic momenta of three fragments (turquoise arrows) together with the generated \cite{Fehre21} \textit{molecular} coordinate system at the instant of photoionization (black axes  $x^\prime,y^\prime$, and $z^\prime$).}\label{fig:MF}
\end{figure}

In order to determine the molecular orientation in spase, three momentum vectors are required and thus only dissociation channels yielding at least three fragments can be considered. Here, we made use of our previous result \cite{Fehre21} and chose so-called \textit{incomplete} breakup channels: $C_3H_6O +\hbar\omega \to C_2H_{2,3}^++CH_2^++OH_{2,1}^0+e^-_{ph}+e^-_{Aug}$. We constructed the \textit{fragment} coordinate system as $\vec{Y}=\vec{p}_{C_2H_{2,3}^+} $,  $\vec{Z}=\vec{p}_{C_2H_{2,3}^+} \times \vec{p}_{CH_2^+} $, and $\vec{X}=\vec{Y}\times\vec{Z} $. We discriminate the selected fragmentation channel from the background by gating on a region in the photoion-photoion-coincidence spectrum and on specific intervals of the magnitude of the momenta of the two measured charged fragments and the third uncharged fragment. The momentum vector of the neutral fragment was computed from the measured momenta of the charged particles exploiting momentum conservation. The gates on the momenta were set to:   $45~\mathrm{a.u.}<\lvert\vec{p}_{CH_2^+}\lvert<105~\mathrm{a.u.}$, $20~\mathrm{a.u.}<\lvert\vec{p}_{C_2H_{2,3}^+}\lvert<110~\mathrm{a.u.}$,  and $20~\mathrm{a.u.}<\lvert\vec{p}_{OH_{2,1}^0}\lvert$. In addition, the condition on the minimum momentum of the neutral fragment ensured that the measured  momenta of charged fragments are not colinear and thus define a plane. For  fragments with several atoms, their center of charge  does not necessarily coincide with their center of mass, which leads to the breakdown of the axial-recoil approximation \cite{Zare72}. Therefore, the \textit{fragment} coordinate system, which is build on their detected asymptotic momenta, is slightly rotated with respect to the \textit{molecular} coordinate system at the instant of photoionization. Details on finding such \textit{molecular} coordinate system with the help of polarization-insensitive MFPADs can be found in our previous work \cite{Fehre21}. Both, the \textit{fragment} ($X,Y,Z$) and the \textit{molecular} ($x^\prime,y^\prime,z^\prime$)  coordinate systems are illustrated in Fig.~\ref{fig:MF}.

\begin{figure*}
\centering
\includegraphics[width=0.95\textwidth]{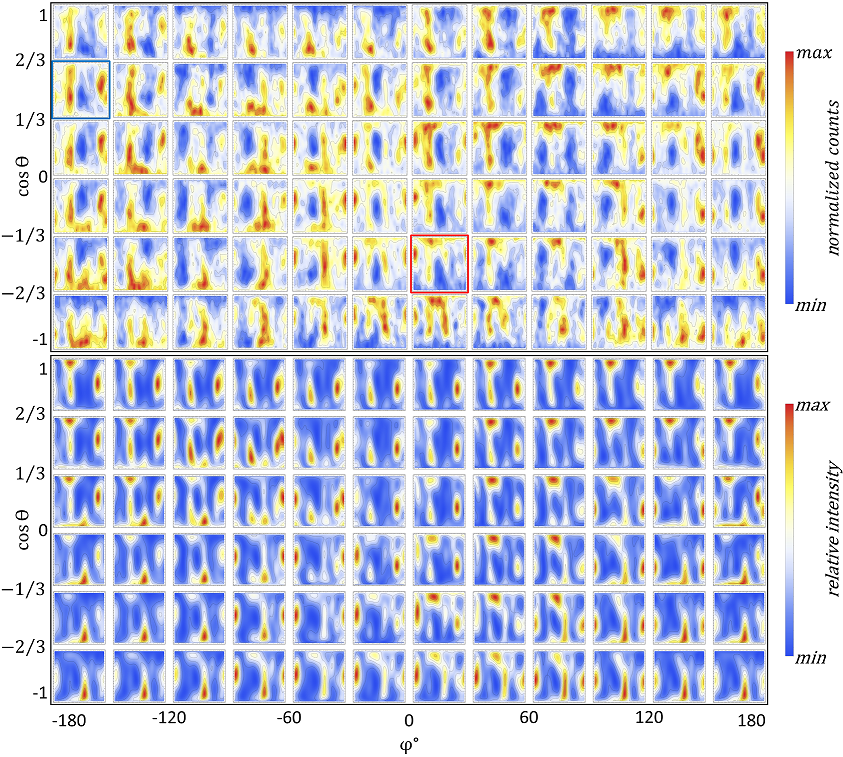}
\caption{Experimental (upper panels) and theoretical (lower panels) fourfold differential photoelectron emission distributions as functions of the photoelectron emission direction and the direction from which the RCP light hits the S-methyloxirane. Both directions are defined by spherical angles $(\theta,\varphi)$ given in the \textit{fragment} ($X,Y,Z$) coordinate system as shown in Fig.~\ref{fig:MF}. The location of each of the 72 patterns indicates the direction from which the light hits the molecule. In the electric-dipole approximation, a reversal of the direction of light in the \textit{fragment} system is equivalent to a swap of the light helicity. Opposite directions  are linked via transformations  $\cos\theta \to - \cos\theta$ and $\varphi \to \varphi+180^\circ$. The positions of two panels highlighted by blue and red squares, which are shown in Fig.~\ref{fig:LCPRCP} on an enlarged scale, obey those transformations. Interchanging the two enantiomers mirrors the coordinate system via reflection at $\cos\theta =0$ for both, the position of the panel and within each panel.}\label{fig:MFPADs}
\end{figure*}

Our measured photoelectron emission distributions are depicted in the upper panels of Fig.~\ref{fig:MFPADs} for different  orientations of S-methyloxirane  with respect to the direction of propagation of the RCP light. For a better representation, the  distributions are shown as functions of the two emission angles ($\theta,\varphi$) in the \textit{fragment} ($X,Y,Z$) coordinate system (see Fig.~\ref{fig:MF}). The light's propagation direction was also transformed in the \textit{fragment} coordinates and is given by the same spherical angles. In order to increase statistics, the orientations from which the light hits the molecule were binned in 72 subsets in constant steps of $\Delta\cos\theta=\frac{1}{3}$ and $\Delta \varphi = 30^\circ$. Here, the position of each panel in the figure indicates the selected light propagation direction in the \textit{fragment} coordinate system. A comparison of this representation for RCP with that for LCP light (not shown for brevity) confirms the validity of the electric-dipole approximation. In this approximation, the light's electric-field vector is fully characterized by its sense of rotation \cite{Pier20}. An inversion of the light propagation direction in the \textit{fragment} frame is, therefore, equivalent to an inversion of the light's helicity. In Fig.~\ref{fig:MFPADs}, such an inversion of the light propagation direction corresponds to a selection of another panel, which is shifted as $\varphi \to \varphi+180^\circ$ and mirrored by $\cos\theta \to - \cos\theta$. Figure~\ref{fig:MFPADs} illustrates a great sensitivity of the measured photoelectron emission distributions to the molecular orientation. This observation is fully confirmed by our \textit{ab initio} calculations (see lower panels in the figure), performed by the single center (SC) method and code \cite{SC1,SC2,SC3}. More details of the calculations can be found in our previous works \cite{Tia17,Hartmann19,Fehre21} on this molecule.

\begin{figure}
\centering
\includegraphics[width=0.45\textwidth]{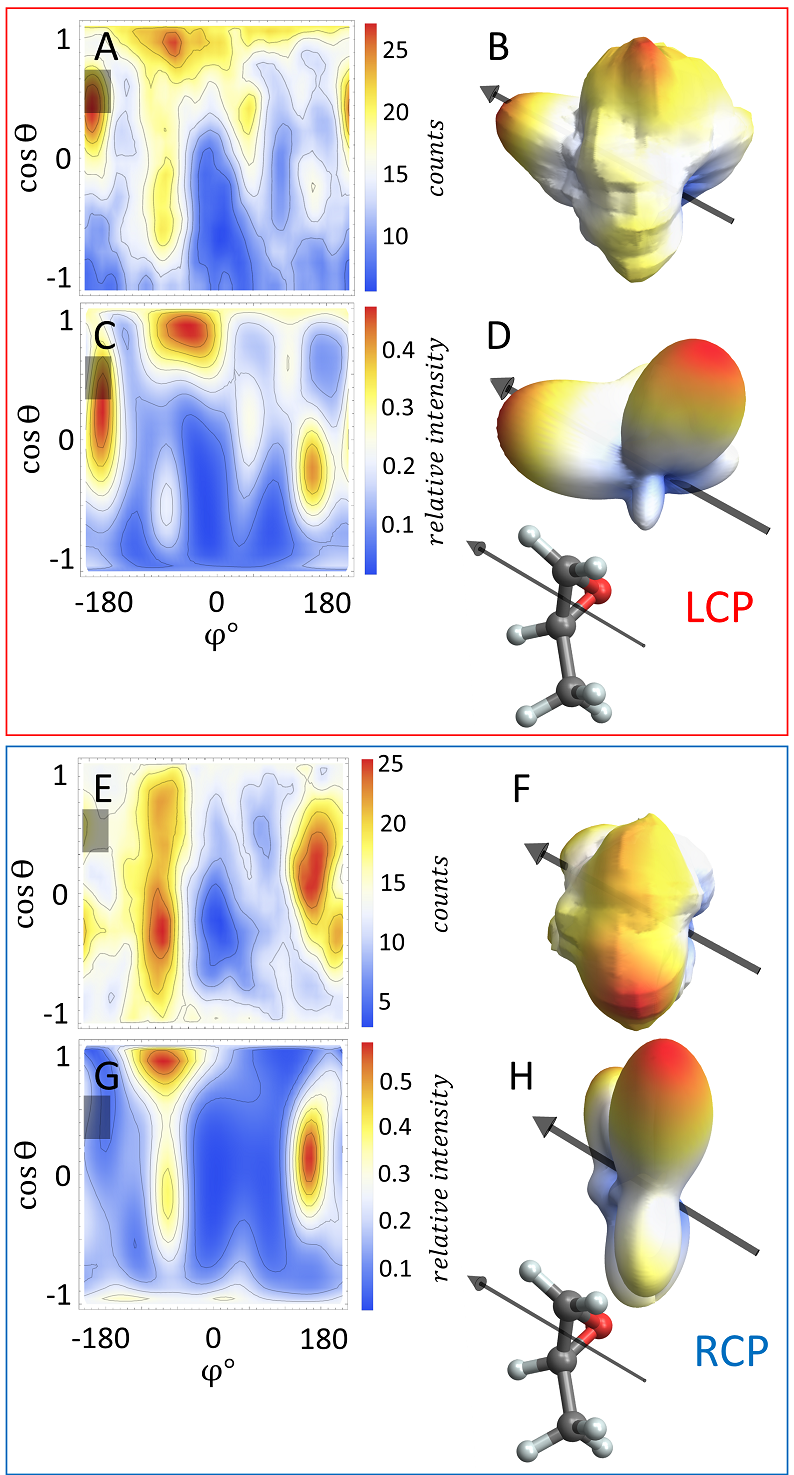}
\caption{Measured (A, B, E, and F) and computed (C, D, G, and H) photoelectron emission distributions in the \textit{fragment} coordinate system for S-methyloxirane and LCP (A--D) or RCP (E--H) light under the condition that the light impinges from the direction indicated by the transparent grey squares (in A, C, E, and G) or grey arrows  (in B, D, F, and H). The molecular  orientation is also visualized in the insets. In panels B, D, F, and H, which  depict the three-dimensional representation of the data from panels A, C, E, and G, respectively, the distance from the origin and the colour code indicate the number of emitted photoelectrons.}\label{fig:LCPRCP}
\end{figure}

A similar symmetry consideration links  our measurements and calculations for S-methyloxirane, shown in Fig.~\ref{fig:MFPADs}, to those for R-methyloxirane (not shown here for brevity). The fragment momenta define the turquoise plane in Fig.~\ref{fig:MF} and thus cannot determine the handedness of the coordinate system. Therefore, the reflection of the \textit{fragment} coordinates at $\cos\theta=0$  mediates switching between two enantiomers. This mirroring operation  $\cos\theta \to - \cos\theta$ ought to be performed for both, the position of each panel in Fig.~\ref{fig:MFPADs} (represents the light propagation direction) and the emission direction of the photoelectron in each panel. The present experimental results on R-methyloxirane confirm this  relationship.  Figure~\ref{fig:LCPRCP} illustrates the  sensitivity of the photoelectron emission pattern obtained at a given molecular orientation to the helicity of the ionizing light. The selected molecular orientation is visualized in the insets. The experimental and theoretical emission distributions are depicted in two representations: as in Fig.~\ref{fig:MFPADs} by the colour maps and also via the three-dimensional polar representation (see caption of the figure for details). Such distributions can be understood as a three-dimensional diffraction pattern of the photoelectron wave, which originates at the oxygen atom and is multiply scattered on the chiral potential of methyloxirane. As one can see from this figure, being imprinted onto the emitted photoelectron waves, the helicity of the ionizing light influences the emission  distributions dramatically. Both, experiment and theory demonstrate  rich structures and sizable contrasts between minima and maxima in the distributions, which are the main  prerequisites for a strong circular dichroism effect.

\begin{figure}
\centering
\includegraphics[width=0.45\textwidth]{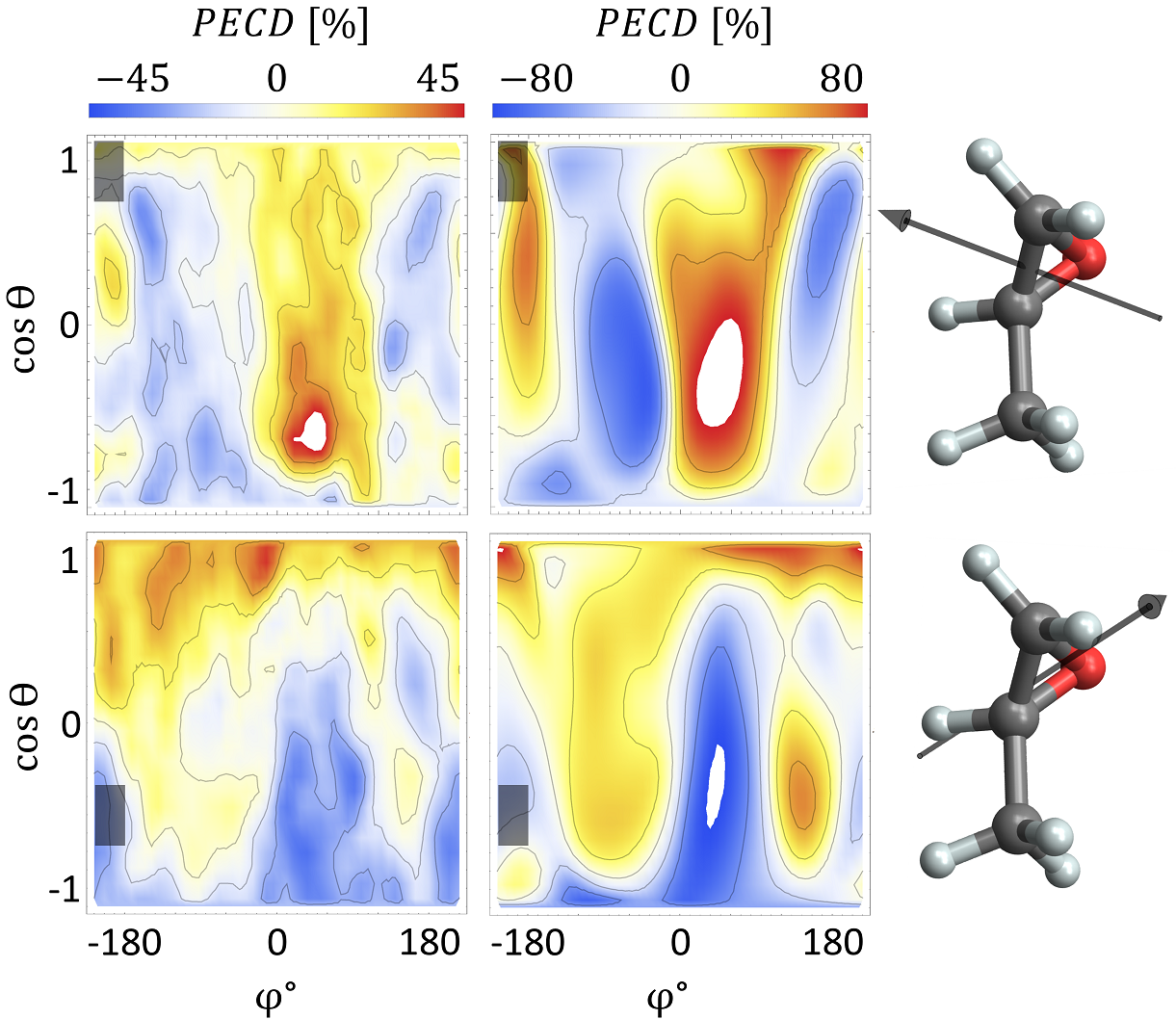}
\caption{Measured (left panels) and computed (right  panels) fourfold differential PECDs for S-methyloxirane as functions of the two photoelectron emission angles for two different molecular orientations, which  are indicated by transparent grey squares and also by grey arrows in the respective insets. White spots highlight photoemission directions at which the measured and computed PECDs exceed the presently chosen upper or lower limits of the respective asymmetry scales.}\label{fig:PECD}
\end{figure}

We finally discuss the fourfold differential PECD of fixed-in-space S-methyloxirane, which is obtained as the normalized difference of the photoelectron emission distributions recorded for LCP and RCP light (indicated by $I_{+}$ and $I_{-}$), respectively: $PECD(\theta,\varphi)=\frac{I_{+}(\theta,\varphi)-I_{-}(\theta,\varphi)}{I_{+}(\theta,\varphi)+I_{-}(\theta,\varphi)}$ \cite{Hergenhahn04,Tia17}. The complete data set can be found in Fig.~S1 of the Supplemental Material \cite{SupplMat}, where the present experimental and theoretical results are depicted in a way similar to that of Fig.~\ref{fig:MFPADs}. For the sake of brevity, we demonstrate here the fourfold differential PECD only for two selected  molecular orientations (see Fig.~\ref{fig:PECD} with the respective insets visualizing the chosen orientations). As one can see from Fig.~\ref{fig:PECD}, for some  photoelectron emission directions, the  measured fourfold differential PECD exceeds 50\%. The measured PECD landscapes and signs are reproduced by our theory in great detail. However, the contrast of the computed PECD is somewhat larger than that of the measured one, well above 80\%. This difference has a twofold origin. On the one hand, our fixed-nuclei calculations overestimate the chiral response, which is, in reality, washed out by the accompanying nuclear dynamics. On the other hand, the experiment underestimates the chiral response. Since an  incomplete breakup channel is used in the analysis, momentum conservation allows to access the momentum of the neutral fragment. This measurement scheme is, however, more prone to noise. In addition, variations in the fragmentation dynamics for the combined breakups with a proton attached to different fragments  cause uncertainties in the determination of the \textit{fragment} coordinate system.

In conclusion, we experimentally and theoretically studied the sensitivity of the O 1s-photoelectron emission pattern of the chiral  methyloxirane molecule to the helicity of circularly polarized light. We examined different orientations of the molecule with respect to the light propagation direction. Our measurements demonstrate that for distinct molecular orientations and photoelectron emission directions, the PECD can be enhanced above 50\%. Taking into account that the measured and computed PECD strengths are also considerably reduced by the limited resolution in our determination of the molecular orientation  (i.e., the binning of only 72 orientations in constant steps of $\Delta\cos\theta=\frac{1}{3}$ and $\Delta \varphi = 30^\circ$), our results fully support the theoretical prediction of Ref.~\cite{Tia17} that the differential PECD of a fixed-in-space chiral molecule can reach 100\%. Such a strong asymmetry may allow for a chiral recognition of gaseous sample molecules with an unprecedented sensitivity.

\begin{acknowledgments}
This work was funded by the Deutsche Forschungsgemeinschaft (DFG) -- Project No. 328961117 -- SFB 1319 ELCH (Extreme light for sensing and driving molecular chirality). K.F. and A.H. acknowledge support by the German National Merit Foundation. M.S.S. thanks the Adolf-Messer Foundation for financial support. H.F. and K.U. acknowledge the XFEL Priority Strategy Program of MEXT, the Research Program of "Dynamic Alliance for Open Innovations Bridging Human, Environment and Materials", and the IMRAM project for support. We acknowledge synchrotron SOLEIL (Saint-Aubin, France) for the provision of experimental facilities at the beamline SEXTANTS. We thank the staff of SOLEIL for running the facility and providing beamtimes under projects 20140056 and 20141178 and especially beamline SEXTANTS for their excellent support. We thank A. Czasch und O. Jagutzki from Roentdek GmbH for support with the detectors.
\end{acknowledgments}

The experiment was conceived by M.S.S. and R.D. The experiment was prepared and carried out by S.G., G.K., S.E., F.T., J.R., A.Har., D.T., C.J., G.N., M.P., S.Z., F.W., M.W., M.K., M.H., L.Ph.H.S., A.K., A.Han., L.B.L., A.E., R.B., H.F., K.U., H.S.-B., J.B.W., T.J., and M.S.S. Data analysis was performed by K.F. and M.S.S. Theoretical calculations were performed by N.M.N. and P.V.D. Initial draft was created by K.F. and P.V.D. All authors discussed the results and commented on the manuscript.

\begin{figure*}
\includegraphics[width=0.95\textwidth]{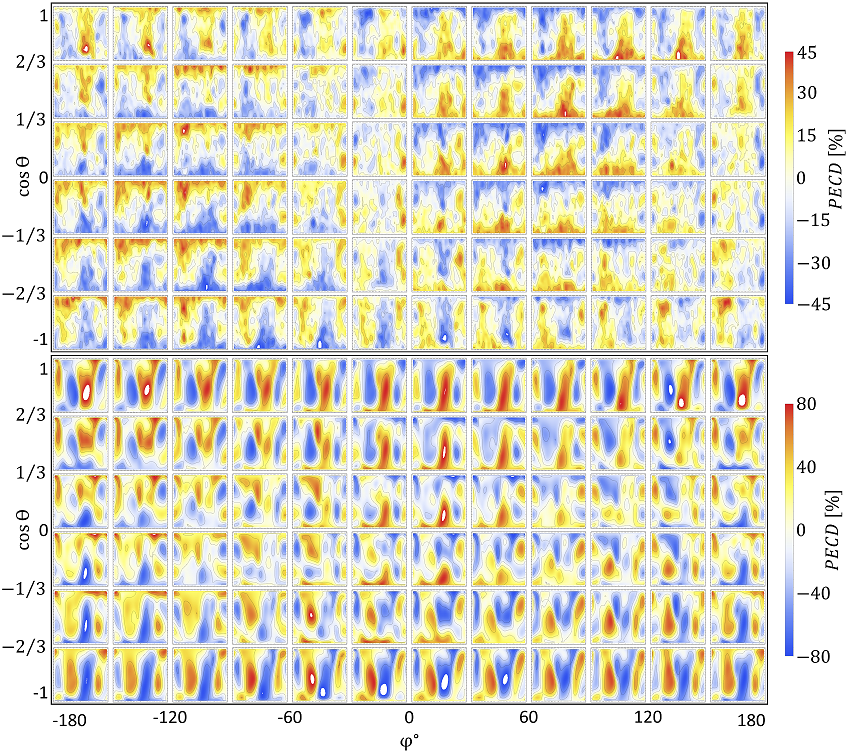}
\justify{FIG. S1: The complete account of the experimental (upper panels) and theoretical (lower panels) fourfold differential PECDs for S-methyloxirane as functions of the two photoelectron emission angles and the two molecular orientation angles (see caption of Fig.~\ref{fig:MFPADs} of the main manuscript for details on the data representation). For some molecular orientations,  white spots highlight photoemission directions at which the measured and computed PECDs exceed the presently chosen upper or lower limits of the respective asymmetry scales.}
\end{figure*}

\end{document}